\documentclass[aps,pra,twocolumn,showpacs,superscriptaddress]{revtex4-1}

\usepackage{amsfonts,amssymb,amsmath}
\usepackage{mathtools}
\usepackage[]{graphics,graphicx,epsfig}
\usepackage{amsthm,multirow}
\usepackage{color}
\usepackage{natbib}

\begin{document}

\title{Single-photon quantum nonlocality: Violation of the Clauser-Horne-Shimony-Holt inequality using feasible measurement setups}

\author{Su-Yong Lee}
\email{papercrane79@gmail.com}
\affiliation{School of Computational Sciences, Korea Institute for Advanced Study, Hoegi-ro 85, Dongdaemun-gu, Seoul 02455, Korea}

\author{Jiyong Park}
\affiliation{Department of Physics, Texas A$\&$M University at Qatar, P.O. Box 23874, Doha, Qatar}

\author{Jaewan Kim}
\affiliation{School of Computational Sciences, Korea Institute for Advanced Study, Hoegi-ro 85, Dongdaemun-gu, Seoul 02455, Korea}

\author{Changsuk Noh}
\email{changsuk@kias.re.kr}
\affiliation{School of Physics, Korea Institute for Advanced Study, Hoegi-ro 85, Dongdaemun-gu, Seoul 02455, Korea}

\date{\today}

\begin{abstract}
We investigate quantum nonlocality of a single-photon entangled state under feasible measurement techniques consisting of on-off and homodyne detections along with unitary operations of displacement and squeezing. We test for a potential violation of the Clauser-Horne-Shimony-Holt (CHSH) inequality, in which each of the bipartite party has a freedom to choose between 2 measurement settings, each measurement yielding a binary outcome. We find that single-photon quantum nonlocality can be detected when two or less of the 4 total measurements are carried out by homodyne detection. The largest violation of the CHSH inequality is obtained when all four measurements are squeezed-and-displaced on-off detections. We test robustness of violations against imperfections in on-off detectors and single-photon sources, finding that the squeezed-and-displaced measurement schemes perform better than the displacement-only measurement schemes.
\end{abstract}

\maketitle

\section{Introduction} 

Ever since the monumental discovery by Bell that quantum nonlocality (QN) can be tested experimentally \cite{Bell1964}, quantum entanglement has captured interest of many scientists, ultimately leading to the development of  the field of quantum information and computation \cite{NielsenChuang2011}.

Traditionally, quantum nonlocality has been investigated in systems with two or more particles, following the pioneering exposition by Einstein, Podolsky, and Rosen \cite{EinsteinRosen1935}, but in the early 90's it has been realized that single-photon states are also capable of exhibiting QN \cite{TanCollett1991,Hardy1994,OliverStroud1989}. This came as a surprise, because single particle states show no signs of entanglement when expressed as a simple superposition of wavefunctions, and initially there were doubts as to whether single-particle quantum nonlocality is genuine. Subsequent theoretical \cite{BanaszekWodkiewicz1999,LeeKim2000,BjorkSanchez-Soto2001,vanEnk2005,DunninghamVedral2007,CooperDunningham2008,BraskBrunner2013} and experimental \cite{LombardiDe-Martini2002,SciarrinoDe-Martini2002,BabichevLvovsky2003,BabichevLvovsky2004,BabichevLvovsky2004a,HessmoBjork2004,ChoiKimble2008} investigations confirmed that single-particle QN is indeed genuine and clarified that the entanglement lies between the spatial modes, rather than between the particles (a clear and concise summary can be found in Ref.~\cite{vanEnk2005}). Furthermore, it has been shown that single-photon entangled states are useful for quantum information processing tasks \cite{KnillMilburn2001,LeeKim2003,SalartZbinden2010,SangouardGisin2011}. 

Direct experimental verifications of single-photon quantum nonlocality (SPQN) \cite{BabichevLvovsky2004a,HessmoBjork2004} involved post-selection and therefore were not decisive. For this reason, the search for experimentally feasible tests of SPQN still holds a merit. In fact, a direct verification of a less restrictive nonlocality measure called EPR-steering \cite{WisemanDoherty2007,JonesWiseman2011} has been demonstrated very recently \cite{FuwaFurusawa2015,Thew2016}. Furthermore, a robust experimentally-feasible SPQN test will be useful for checking the security of quantum key distribution protocols that use vacuum-one-photon qubits \cite{LeeKim2003,LeeBergou2009}, as well as for making a non-trivial self-testing statement \cite{Jed2016}.

In this work, we investigate the feasibility of testing SPQN using widely-employed experimental measurement techniques of on-off detection and homodyne measurements. We test the violation of the Clauser-Horne-Shimony-Holt (CHSH) \cite{ClauserHolt1969} inequality using these detection techniques, further allowing unitary squeezing and displacement operations to assist the measurements. The latter are chosen because they have been demonstrated in many quantum optics experiments. Because homodyne measurements are generally more efficient than photon counting, we pay close attention to whether they can be used to detect QN. A survey of possible combinations of Gaussian-assisted on-off and homodyne detections reveals that on-off detections are necessary to detect QN, while the addition of homodyne measurements is detrimental. No violation of the CHSH inequality is observed when three or more homodyne measurements are involved. 

Within the scheme, the maximum violation is achieved when all measurements are squeezed-and-displaced on-off detections, giving a larger violation than the previously reported displaced on-off detection scheme. We show that additional squeezing operations provide robustness against imperfections in the single-photon source and also against detection losses. 
The remainder of this paper is organized as follows. Sect.~II describes our scheme for testing QN, including measurement strategies and analytic formulae involved.
Whether QN can be observed in various scenarios within the scheme is reported in Sect.~III and robustness of violations against imperfections are treated in Sect.~IV. We conclude with a discussion in Sect.~V.

\section{Schemes for testing quantum nonlocality}
As mentioned above, we will focus on two experimentally feasible measurement techniques, on-off detection and homodyne measurement, assisted by two unitary (Gaussian) operations, displacement and squeezing. The single-photon state is chosen to be
\begin{align}
\label{eq1}
|\psi\rangle = \frac{1}{\sqrt{2}}\left(|1\rangle_A|0\rangle_B+|0\rangle_A|1\rangle_B\right).
\end{align}
The Bell-type scheme for detecting SPQN is illustrated in Fig.~1: two spatially separated parties Alice and Bob perform a set of possible measurements $M_A$ and $M_B$, respectively. We consider two-types of measurements denoted $M_1$ and $M_2$, which we take to be Gaussian operation-assisted on-off and homodyne measurements respectively. We focus on the case in which each party has two measurement settings with binary outcomes. 
\begin{figure}[ht]
\centerline{\scalebox{0.35}{\includegraphics[angle=-90]{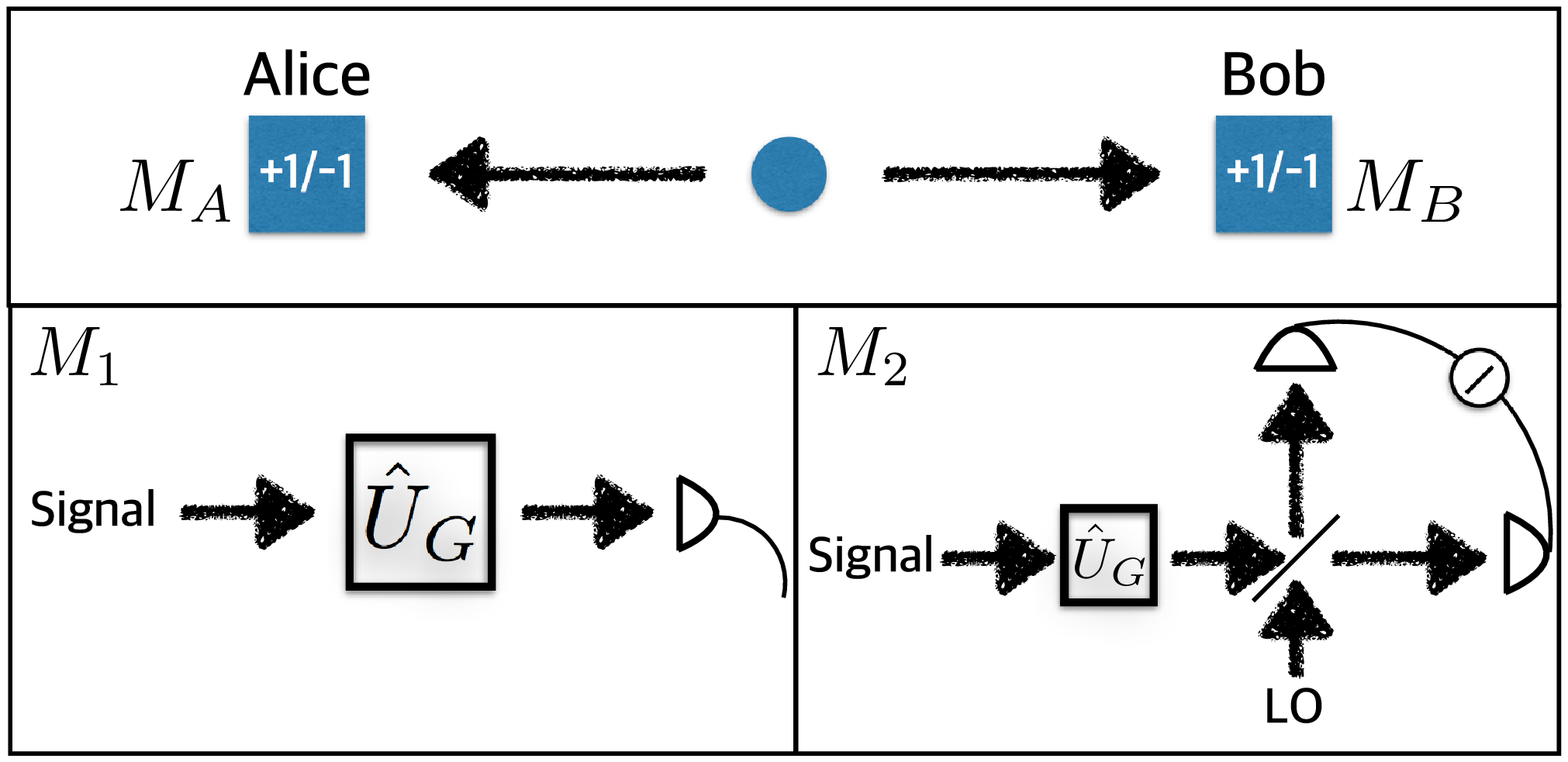}}}
\vspace{-1.5in}
\caption{Schemes for testing single-photon quantum nonlocality using feasible measurement setups. $M_A$ ($M_B$) denote possible measurement settings of Alice (Bob), and $+1/-1$ denote their measurement outcomes.
$M_{A,B}$ can be either an on-off ($M_1$) or a homodyne detection ($M_2$), assisted by Gaussian-operations. $\hat{U}_G$ denotes a Gaussian operation comprised of displacement and/or squeezing, and LO denotes a local oscillator (strong laser field).}
\label{fig:fig1}
\end{figure} 

Let us denote Alice's (Bob's) measurements by $A_m$ ($B_m$) with eigenvalues $\in \{+1,-1\}$ and the corresponding measurement settings $a_m$ $(b_m) \in \{ 1,2\}$.
The `amount' of QN can be measured by the degree of violation of the CHSH inequality \cite{ClauserHolt1969}
\begin{eqnarray}
\label{CHSH}
S=|E(A_1 B_1)+E(A_1 B_2)+E(A_2 B_1)-E(A_2 B_2) |\leq 2,\nonumber\\
\end{eqnarray}
where $E(A_m B_n)$ is the expectation value of the operator product $A_m B_n$, that is, $\langle A_m B_n \rangle$.

\subsection{Measurement strategies}
An on-off detector measures the presence (on) or absence (off) of photons and is generally not able to resolve the number of photons. To test the CHSH inequality, we allocate a value 1 to the on (click) events and -1 to the off (no-click) events.  Accordingly, the on-off detection operator can be written as $\hat{O}\equiv\sum^{\infty}_{n=1}|n\rangle\langle n|-|0\rangle\langle 0|=\hat{I}-2|0\rangle\langle 0|$. The effects of Gaussian operations can be taken into account by changing the operator to $\hat{O}' = \hat{U}^\dagger_G\hat{O}\hat{U}_G$

A (balanced) homodyne detector measures the intensity difference between the two output modes produced by combining the signal and local oscillator fields using a 50:50 beam splitter (see Fig.~\ref{fig:fig1}). The resulting photoelectric current is related to the expectation value of a field quadrature operator via the result $\langle\hat{X}_{\theta}\rangle=\Delta I/(\sqrt{2}|\alpha_{LO}|)$, where $\Delta I$ is the intensity difference, $\alpha_{LO}$ is the amplitude of the local oscillator field (a laser field), and $\theta$ is the phase of the local oscillator \cite{Leonhardt1997}.
To adopt this detection technique in the CHSH scheme, one must subject an outcome to a binary binning process: when a measurement outcome falls within a certain region, $Z_+$, on the real line we assign a value $+1$, whereas if it falls in the rest of the real line, $Z_-$, we assign a value $-1$. We can thus define a homodyne-binning operator $\hat{X}_{\theta}\equiv (\int_{Z_+} -\int_{Z_-} )|x_{\theta}\rangle\langle x_{\theta}|dx_{\theta}$. For pure homodyne measurements, we adopt the center-binning convention $Z_+=[-z,z]$ ($Z_-=(-\infty,\infty)-Z_+$) \cite{DistanteAndersen2013}. 
Then a general Gaussian operation shifts, rescales, and rotates the profile.

To see this, we explictly look into how a homodyne distribution $h_{\rho} ( x_{\theta} )$ for a quantum state $\rho$ changes under a general Gaussian operation $\hat{U}_{G} = \hat{D} (-\alpha) \hat{S} (\xi)$, where $-\alpha$ is a complex displacement amplitude and $\xi = r e^{i\varphi}$ is a complex squeezing parameter. Using the relation between the characteristic function $C_{\rho} ( \lambda ) \equiv \mathrm{tr} [ \rho \hat{D} ( \lambda ) ]$ and homodyne distribution \cite{RiskenVogel1989},
	\begin{equation}
		 h_{\rho} ( x_{\theta} ) = \frac{1}{\sqrt{2} \pi} \int_{-\infty}^{\infty} dk e^{i\sqrt{2}kx_{\theta}} C_{\rho} ( \lambda = -ike^{i\theta} ),
	\end{equation}
and $C_{\hat{U}_{G} \rho \hat{U}_{G}^{\dag}} ( \lambda ) = C_{\rho} ( \lambda \cosh r + \lambda^{*} e^{i\varphi} \sinh r ) e^{\lambda^{*} \alpha - \lambda \alpha^{*}}$ \cite{BarnettRadmore1997}, we obtain
	\begin{equation}
		h_{\hat{U}_{G} \rho \hat{U}_{G}^{\dag}} ( x_{\theta} ) = \frac{1}{s} h ( \frac{x_{\phi} + \delta}{s} ),
	\end{equation}
where
	\begin{align}
		\delta & = \frac{\alpha e^{-i\theta} + \alpha^{*} e^{i \theta}}{\sqrt{2}}, \nonumber \\
		s & = \sqrt{\cosh(2r)-\sinh(2r)\cos(\varphi-2\theta)}, \nonumber \\
		\phi & = \theta - \arctan \bigg[ \frac{\sinh r \sin ( \varphi - 2\theta )}{\cosh r - \sinh r \sin ( \varphi - 2\theta )} \bigg].
	\end{align}
The displacement shifts the center of the profile, and the squeezing operation rotates and rescales the profile. It yields
	\begin{equation}
		\mathrm{tr} ( \hat{U}_{G} \rho \hat{U}_{G}^{\dag} \hat{X}_{\theta} ) = \bigg( 2 \int_{(-z+\delta)/s}^{(z+\delta)/s} - \int_{-\infty}^{\infty} \bigg) dx_{\phi} h_{\rho} ( x_{\phi} ),
	\end{equation}
 indicating that we can incorporate the effects of the Gaussian operation by setting a proper phase angle and modifying the integration region.

\subsection{Expectation values of the correlation operators}
To evaluate the correlation function $S$ in Eq.~(\ref{CHSH}), one needs to calculate the four expectation values 
$E(A_m B_n)$, the functional form of which will be provided in this subsection. There are four-types of correlation functions depending on whether $A_m$ and $B_n$ correspond to homodyne-binning or on-off measurements.  Following the notations in the previous subsection, the four possible correlation functions can be written as
\begin{subequations}
\label{E}
\begin{align}
E(A_1 B_1)&=\langle \psi |\hat{U}^{\dag a}_G\hat{O}_a\hat{U}^a_G \otimes \hat{U}^{\dag b}_G\hat{O}_b\hat{U}^b_G|\psi\rangle \nonumber\\
&\equiv E_{11}(\alpha,\beta,\xi_{a},\xi_{b}),\\
E(A_1 B_2)&=\langle \psi |\hat{U}^{\dag a}_G\hat{O}_a\hat{U}^a_G \otimes \hat{U}^{\dag b}_G\hat{X}^b_{\theta}\hat{U}^b_G|\psi\rangle \nonumber \\ 
&\equiv E_{12}(\alpha,\xi_a,z_3,z_4),\\
E(A_2 B_1)&=E_{12}(\beta,\xi_b,z_1,z_2),\\
E(A_2 B_2)&=\langle \psi |\hat{U}^{\dag a}_G\hat{X}^a_{\theta}\hat{U}^a_G \otimes \hat{U}^{\dag b}_G\hat{X}^b_{\theta}\hat{U}^b_G|\psi\rangle \nonumber \\ 
&\equiv E_{22}(z_1,z_2,z_3,z_4),
\end{align} 
\end{subequations}
where the operators are given by $\hat{U}^{\dag}_G\hat{O}\hat{U}_G=\hat{I}-2\hat{U}_G^\dagger|0\rangle\langle 0|\hat{U}_G$ and $\hat{U}^{\dag}_G\hat{X}_{\theta}\hat{U}_G=2\int^{z_2}_{z_1}dx_{\phi}|x_{\phi}\rangle\langle x_{\phi}|-\hat{I}$, and the input state $|\psi\rangle$ is given in Eq.~(\ref{eq1}).
The subscripts 1 and 2 refer to (Gaussian-assisted) on-off and homodyne measurements respectively. The subscript (superscript) attached to $\xi,~\hat{O}$
($\hat{U}_G,~\hat{X}_{\theta}$) refer to Alice ($a$) and Bob ($b$), while the displacement amplitudes of Alice and Bob are denoted by $\alpha$ and $\beta$ respectively. 

 All results in this work have been obtained by numerically optimizing (maximizing) the CHSH correlation $S$, as a function of squeezing $\xi$, displacement $\alpha$, and integration interval $[z_1,z_2]$ parameters.
 For example, when each party uses both the homodyne and on-off schemes, $S$ is a function of 14 real parameters ($\alpha,\beta,\xi_a,\xi_b,z_1,z_2,\phi_1,z_3,z_4,\phi_2$), which is optimized. 
 
\section{Violation of the CHSH inequality}

 As we have mentioned in the introduction, homodyne measurements enjoy significantly higher efficiencies than on-off detectors in general. It is thus easier to close the detection loophole when a scheme involves more homodyne detectors than on-off detectors. For this reason we start from the case in which all measurements are homodyne and decrease the number of homodyne measurements in subsequent subsections: (A) 4 homodyne measurements, (B) 3 homodyne and 1 on-off, (C) 2 homodyne/2 on-off, (D) 1 homodyne/3 on-off, (E)  4 on-off detections. Our calculation shows that the violation of the CHSH inequality occurs only in cases (C), (D), and (E). That is, when there are 2 or less homodyne measurements involved.  The maximum values are obtained when there are small amounts of displacement and squeezing.

\subsection{4 homodyne measurements}
In this case, Alice and Bob perform homodyne measurements only. The measurement operators are $A_1=M_1(z^a_1,z^a_2,\phi^a)$ and $A_2=M_2(z'^a_1,z'^a_2,\phi '^{a})$ for Alice and $B_1=M_2(z^b_1,z^b_2,\phi^b)$ and $B_2=M_2(z'^b_1,z'^b_2,\phi '^b)$ for Bob.
Numerical optimization indicates that QN cannot be demonstrated using this measurement scheme. In the sign-binning case, ($Z_+ =(-\infty,0]$), the measurement operator reduces to $\hat{X}_\phi  = \sqrt{2/\pi}(\cos\phi \sigma_x + \sin\phi \sigma_y)$, and it is quite simple to see that there can be no violation of the CHSH inequality  (this result was also found independently by some of us in Ref.~\cite{ParkNha2012}) because of the factor $\sqrt{2/\pi}$ \cite{QuintinoCunha2012,MorinSangouard2013}. Our results indicate that a more general `off-center' binning strategy does not change this result.
 
\subsection{3 homodyne measurements}
 We choose Alice as the one who performs the only on-off measurement. Thus the measurement operators are given by $A_1=M_1(\alpha,\xi_a)$ and $A_2=M_2(z^a_1,z^a_2,\phi^a)$ for Alice and $B_1=M_2(z^b_1,z^b_2,\phi^b)$ and $B_2=M_2(z'^b_1,z'^b_2,\phi '^b)$ for Bob. Equations (\ref{E}b) and (\ref{E}d) along with numerical optimization reveal that no violation of the CHSH inequality occurs.
 
\subsection{2 homodyne measurements}

There are two possible scenarios: i) one party uses Gaussian-assisted on-off detection only, while the other party uses homodyne binning strategy only; ii) Each party uses both types of measurements. 
\subsubsection{Case i}
Alice's measurement operators are $A_1=M_1(\alpha,\xi_a),~A_2=M_1(\beta,\xi_b)$, and Bob's measurement operators are
$B_1=M_2(z^b_1,z^b_2,\phi^b),~B_2=M_2(z'^b_1,z'^b_2,\phi '^b)$. 
Equations (\ref{E}b) and (\ref{E}c) along with numerical optimization yield a maximum value of $S\approx 2.126$ when the parameters are set to: $\alpha = -0.815 - i 0.171$, $\xi_a= -0.332e^{i0.413}$, $\beta = -0.155 +i 0.818$, $\xi_b = 0.332e^{i0.374}$, $z^b_1 = -0.139$, $z^b_2 = 5.237$, $\phi^b = -0.589$, $z'^b_1 = 0$, $z'^b_2 = 8.256$,  and $\phi '^b = 0.982$. 

\subsubsection{Case ii}
The measurement operators are given by $A_1=M_1(\alpha,\xi_a)$, $A_2=M_2(z_1^a,z_2^a,\phi^a)$ for Alice and $B_1=M_1(\beta,\xi_b)$, $B_2=M_2(z_1^b,z_2^b,\phi^b)$ for Bob. 
We find a maximum value of $S\approx 2.231$ when the parameters are set to: $\alpha = 0.264+i0.578$, $\xi_a= 0.24e^{-i0.858}$, $z^a_1 = -11.7$, $z^a_2 =0.143$, $\phi^a = -0.55$, $\beta = 0.153 -i 0.617$, $\xi_b = 0.24e^{i0.486}$, $z^b_1 = 0.143$, $z^b_2 = 9.7$,  and $\phi^b = 0.363$.   We note that a similar violation ($\approx 2.25$) has been found for the 2-photon equivalent of our entangled state in Ref.~\cite{CavalcantiScarani2011}, although Gaussian operations were not included. It would be interesting to see how much the violation can be enhanced with the help of displacement and squeezing in that case.

\subsection{1 homodyne measurement}
The measurements are $A_1=M_1(\alpha,\xi_a)$ and $A_2=M_2(z^a_1,z^a_2,\phi^a)$ for Alice and $B_1=M_1(\beta,\xi_{b})$ and $B_2=M_1(\beta',\xi'_{b})$. Equations (\ref{E}a-c) along with numerical optimization yield a maximum value of $S \approx 2.557$ when the parameters are set to: $\alpha = 0.0$, $\xi_a= 0.0$,  $z^a_1 = -11.5$, $z^a_2 =0.0$, $\phi^a = -0.146$, $\beta = -0.344+i0.051$, $\xi_b= -0.099e^{-i0.293}$, $\beta' = 0.344 -i0.151$, and $\xi '_b = -0.099e^{-i0.293}$. 

 \subsection{0 homodyne measurement}
 We first consider the displacement-only scenario. Alice (Bob) can choose between two displacement amplitudes $\alpha$ ($\beta$) and $\alpha'$ ($\beta'$). 
We find violations of the CHSH inequality with the maximum correlation value as large as $2.688$, which is obtained for the parameters $|\alpha|=|\beta|\approx 0.165$, $|\alpha'|=|\beta'|\approx 0.563$,  $\phi_{\alpha}\approx -3.395$, $\phi_{\beta}\approx 2.888$, and $\phi_{\alpha'}=\phi_{\beta'}\approx -0.253$. The same result has been reported in Ref.~\cite{InvernizziBanaszek2005}. 
 
Next, considering local squeezing only, we find no violation of the CHSH inequality. However, using both the displacement and squeezing, we observe a larger violation, which is also more robust against source- and detection-inefficiencies as we show in the next section.  Displacement and squeezing operations have been shown to be helpful for detecting the violation of two-mode squeezed state with on-off detectors in Ref.~\cite{RyuLee2010}, and furthermore single-mode squeezing was shown to be useful for protecting non-Gaussian states from a loss channel \cite{Filip2013,ParkNha2015}. Let us discuss the ideal case first. Using Eqs.~(\ref{E}) and numerical optimization, we find that $S \approx 2.782$ at $\xi_a=\xi_b= 0.032$, $\xi'_a = \xi'_b= 0.243$, $\alpha=\beta\approx  i 0.186$, and $\alpha'=\beta' \approx  - i0.642$. Compared to the case with displacement alone, the maximum correlation value has moved closer to the Tsirelson bound $2\sqrt{2}\approx 2.828$, which is the upper limit of the CHSH correlations allowed by quantum mechanics. The squeezing parameters $r=0.032,0.243$ correspond to $0.28,2.11$ $dB$, respectively, which lie within the experimentally achievable limit.

\section{Robustness against imperfections}
To test the robustness of the violations found in the previous section, we consider two types of imperfections: photon losses in on-off detections and noise in the single-photon source. Imperfections in homodyne measurements are neglected for simplicity. We also test the role of squeezing by comparing the maximum value of the CHSH correlation for both the squeezed-and-displaced and the displacement-only cases. 
 
The effects of losses in on-off detection can be represented by a two-component positive-operator-valued measure $\hat{\Pi}_0=\sum_n(1-\eta)^n|n\rangle\langle n|$ (no click) and  $\hat{\Pi}_1=\hat{1}-\hat{\Pi}_0$ (click), where $\eta$ is the detection efficiency \cite{Mogilevtsev1998,RossiParis2004}. To describe the efficiency of the single-photon source, we adopt a simple model in which an incoherent mixture of the vacuum component is added: $p|1\rangle\langle 1|+(1-p)|0\rangle\langle 0|$.

\subsubsection{2 homodyne measurements}

\begin{figure}[ht]
\centerline{\scalebox{0.31}{\includegraphics[angle=270]{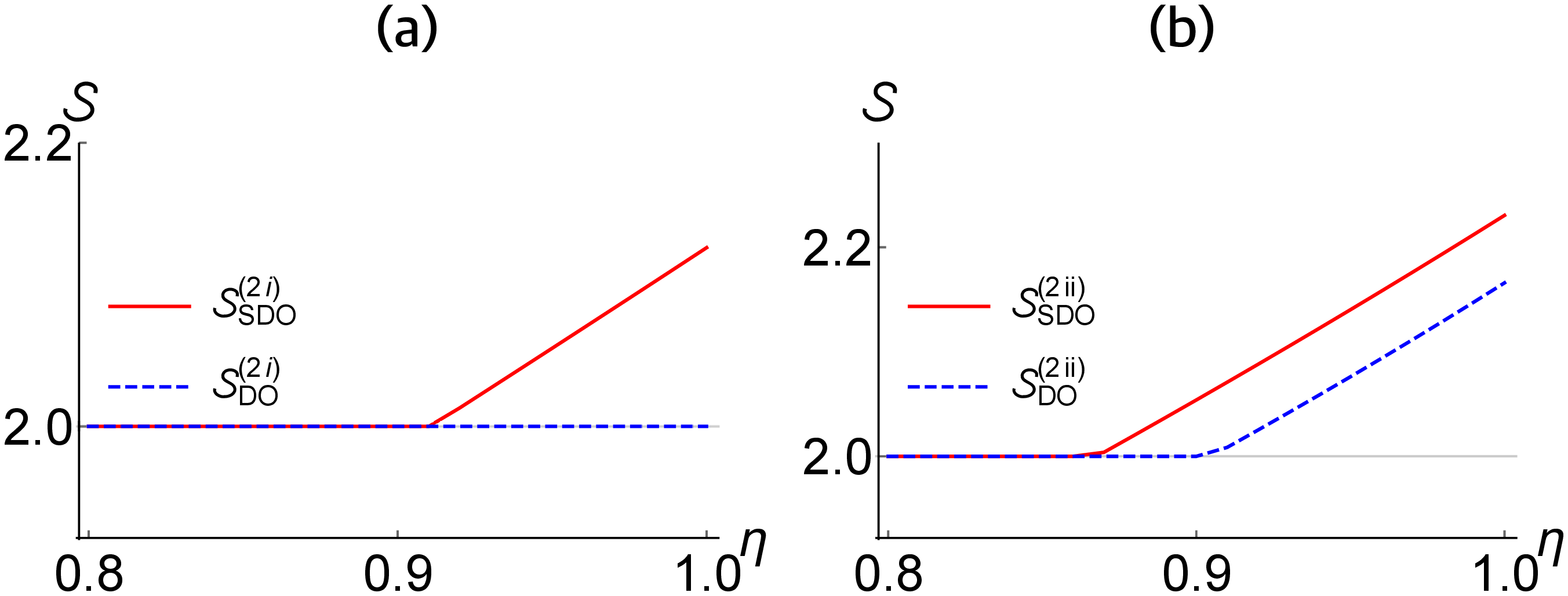}}}
\vspace{-1.1in}
\caption{Maximum values of the CHSH correlation as functions of the detection efficiency for the 2-homodyne case: (a) One party uses two Gaussian-assisted on-off detection, while the other party uses homodyne binning strategies only. (b) Each party uses both types of measurements. The solid (red) curves correspond to the squeezed-and-displaced cases, whereas the dashed (blue) curves are for the displacement-only case. 
} 
\label{fig2}
\end{figure}

Let us denote the maximum value of the CHSH correlation as $S^{(2i)}_{\rm SDO}$ ($S^{(2ii)}_{\rm SDO}$) and $S^{(2i)}_{\rm DO}$ ($S^{(2ii)}_{\rm DO}$), where the subscripts refer to squeezed-and-displaced on-off (SDO) and displaced on-off (DO) schemes. The superscript $i$ ($ii$) refers to the case $i$ ($ii$) in Sec.~III.
Figure \ref{fig2} compares the two cases as functions of the detection efficiency $\eta$, showing that quantum nonlocality of the single-photon entangled state can be verified with a finite detection efficiency. For case $i$ (two homodyne by one party), violations can only be observed when both squeezing and displacement are performed, while for case $ii$, the displacement-only scenario also shows violation although the values are lower than the squeezed-and-displaced scenario.

\begin{figure}[ht]
\centerline{\scalebox{0.31}{\includegraphics[angle=270]{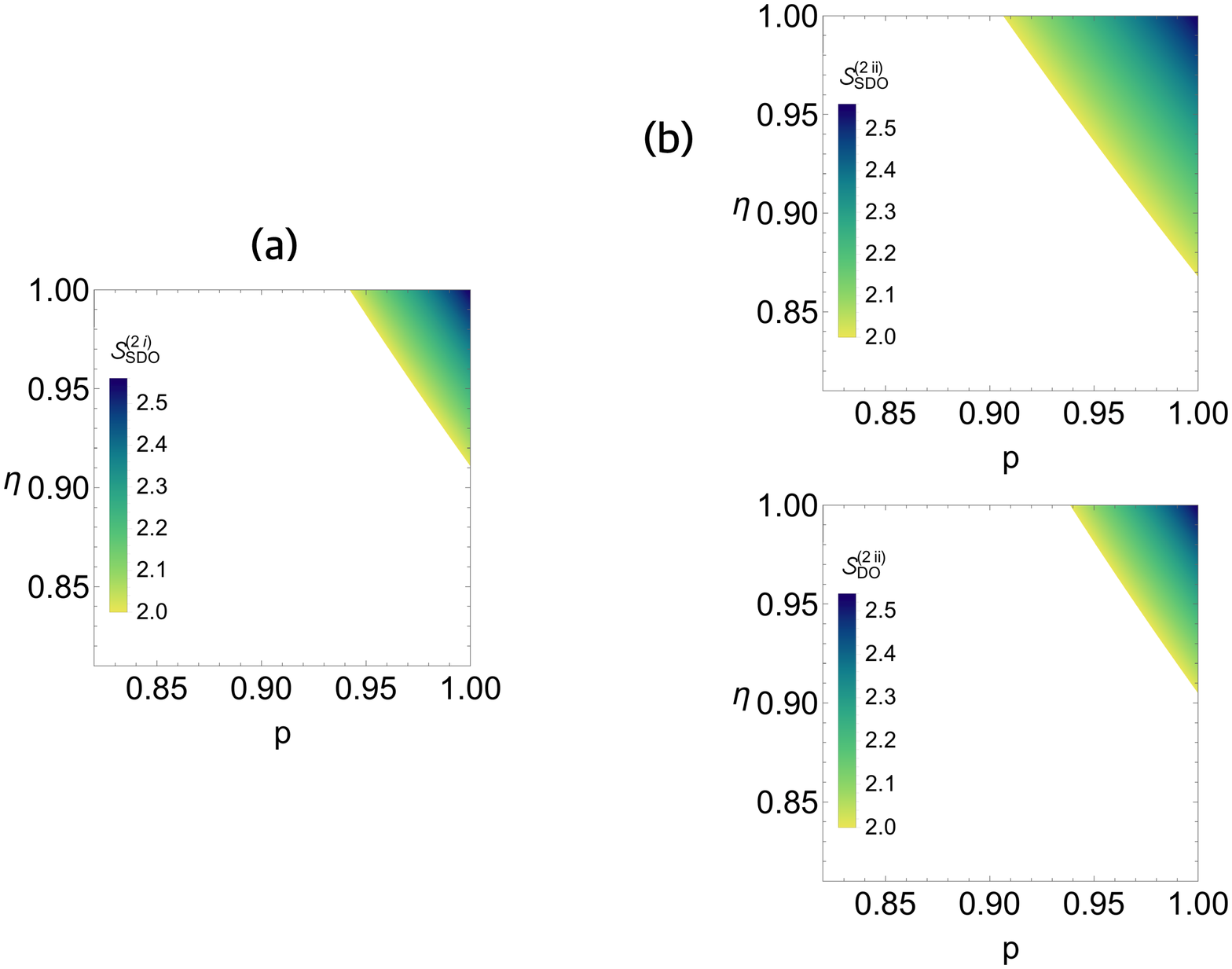}}}
\caption{Maximum values of the CHSH correlation as functions of the detection efficiency $\eta$ and the source efficiency $p$ for the 2-homodyne cases: (a) One party uses two Gaussian-assisted on-off detection, while the other party uses homodyne binning strategies only. (b) Each party uses both types of measurements. Squeezed-and-displaced schemes perform better than the displacement-only schemes in both cases.}
\label{fig3}
\end{figure}

The effects of single-photon source efficiency $p$ and the detection efficiency are depicted in Fig.~\ref{fig3}. The lower bounds for observing quantum nonlocality are located at $\eta\approx 0.91$ and $p\approx 0.942$ for $S^{(2i)}_{\rm SDO}$, at $\eta\approx 0.905$ and $p\approx 0.940$ for $S^{(2ii)}_{\rm DO}$, and at $\eta\approx 0.870$ and $p\approx 0.908$ for $S^{(2ii)}_{\rm SDO}$.

\subsubsection{1 homodyne measurement}
\begin{figure}[ht]
\centerline{\scalebox{0.6}{\includegraphics[angle=0]{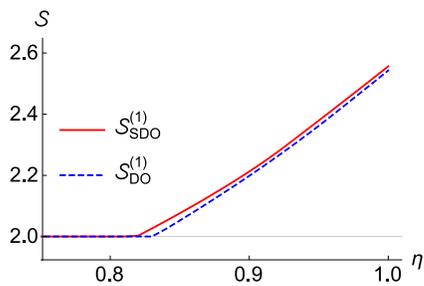}}}
\caption{Maximum values of the CHSH correlation as a function of the detection efficiency for the 1-homodyne case. The solid (red) curve is for the squeezed-and-displaced on-off detection scheme, whereas the dashed (blue) curve is for the displaced on-off detection scheme. The squeezed-and-displaced scheme performs slightly better than the displacement-only scheme.}
\label{fig4}
\end{figure}

\begin{figure}[ht]
\centerline{\scalebox{0.36}{\includegraphics[angle=0]{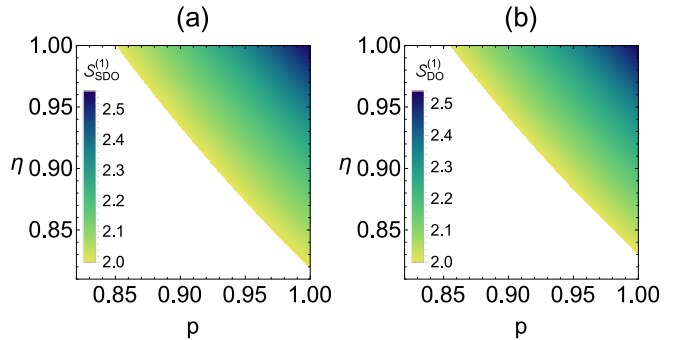}}}
\caption{Maximum values of the CHSH correlation as functions of the detection- and source-efficiencies for the 1-homodyne case: (a) the squeezed-and-displaced on-off detection scheme and (b) the displaced on-off detection scheme. The squeezed-and-displaced scheme has a slightly lower bound for both efficiencies than the displacement-only scheme.}
\label{fig5}
\end{figure}

The maximum values of the CHSH correlation are denoted $S^{(1)}_{\rm SDO}$ and $S^{(1)}_{\rm DO}$. Figure \ref{fig4} compares the two as functions of the detection efficiency $\eta$, showing a similar behaviour to the 2-homodyne case: higher violation is observed with extra squeezing.
 Figure \ref{fig5} plots the maximum values for the two cases as functions of both the detection and source efficiencies. A slightly larger area of violation is observed for the squeezed-and-displaced case.

\subsubsection{0 homodyne measurements}
\begin{figure}[ht]
\centerline{\scalebox{0.6}{\includegraphics[angle=0]{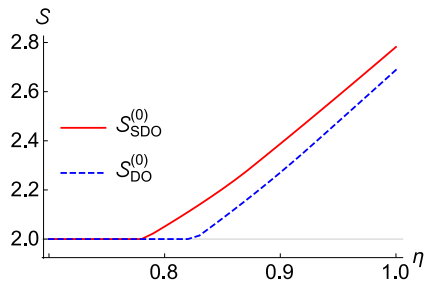}}}
\caption{Maximum values of the CHSH correlation as a function of the detection efficiency for the 0-homodyne case. The solid (red) curve is for the squeezed-and-displaced on-off detection scheme, whereas the dashed (blue) curve is for the displaced on-off detection scheme. The squeezed-and-displaced scheme performs better than the displacement-only scheme.
} 
\label{fig6}
\end{figure}

\begin{figure}[ht]
\centerline{\scalebox{0.36}{\includegraphics[angle=0]{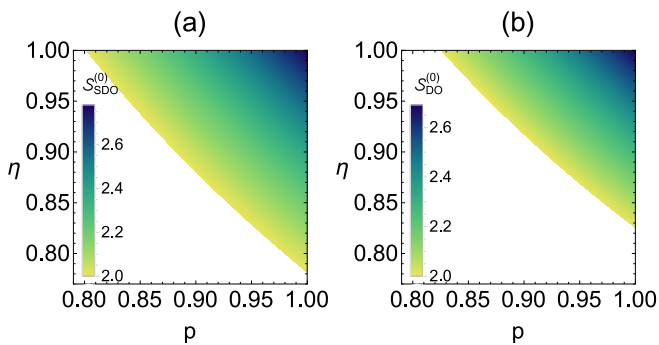}}}
\caption{Maximum values of the CHSH correlation as functions of the detection- and source-efficiencies for the 0-homodyne case: (a) the squeezed-and-displaced on-off detection scheme and (b) the displaced on-off detection scheme. The squeezed-and-displaced scheme has significantly lower bounds for both inefficiencies than the displacement-only scheme.
}
\label{fig7}
\end{figure}

Let us denote the maximum values of the CHSH correlation as $S^{(0)}_{\rm SDO}$ and $S^{(0)}_{\rm DO}$. One more (Gaussian-assisted) on-off detection has been added to the measurement scheme, so one would expect the effect of detection efficiency to be more detrimental. This expectation can be verified by comparing the slopes of the curves in Figs.~\ref{fig4} and \ref{fig6}. The slopes in the 1-homodyne case are indeed shallower. However, all in all, the 0 homodyne-measurement scheme is more robust against imperfections as can be verified by comparing Fig.~\ref{fig5} and Fig.~\ref{fig7}: In the squeezed-and-displaced case, the lower bounds for observing quantum nonlocality are located at $\eta\approx 0.82$ and $p\approx 0.85 $ for the 1-homodyne scheme, and at $\eta\approx 0.78$ and $p\approx 0.8$ for the 0-homodyne scheme. In the displacement only case, the corresponding bounds are located at $\eta\approx 0.83$, $p\approx 0.855$ and $\eta\approx 0.825$, $p\approx 0.83$ (the same bound for $\eta$ was also obtained in Ref. \cite{LeeJaksch2009}, and the bounds for the W-state input were found in Ref. \cite{Brask2012}) .  This behaviour can be understood as a result of higher maximum values of the correlation.

As for the effects of an extra squeezing operation, we see that the bounds have moved from $(\eta \approx 0.825, p\approx 0.83)$ to $(\eta \approx 0.78, p\approx 0.8)$. Compared to the 1-homodyne case, the squeezing adds significant robustness to the testing scheme against source and detector inefficiencies.

\section{Summary and discussion}
We have investigated the feasibility of demonstrating quantum nonlocality of a single-photon entangled state using on-off and homodyne measurements assisted by local Gaussian operations, i.e., displacement and single-mode squeezing operations. 
We have found that the violation of the CHSH inequality can be demonstrated when there are two or less homodyne-measurement involved (out of 4 total measurements). 
The maximum CHSH correlation values are: $S\approx 2.231$ for two homodyne-measurement case, $S\approx 2.557$ for one homodyne-measurement case, and $S\approx 2.782$ for the 0 homodyne-measurement case.
Previously reported maximum value was at $S\approx 2.688$ using on-off measurements with displacement operations only.
Our result shows that the additional single-mode squeezing operations enhance the violation. Table \ref{tab} summarizes our finding for all possible measurements, displaying the maximum achievable CHSH correlations.
\begin{table}[]
\caption{Testing single-photon entangled states using Homodyne Detection (HD) and On-off Detection with displacement (D) operation and single-mode squeezing (S) operation, in CHSH inequality.} 
\begin{tabular}{|c|c|c|}
\hline  Alice ($A_1,~A_2$)~& Bob ($B_1,~B_2$) & CHSH  \\
\hline\hline  HD & HD &$\leq 2$   \\
\hline HD & HD, On-off with D \& S & $\leq 2$  \\
\hline HD & On-off with D  & $\leq 2$   \\
\hline HD & On-off with D \& S & $2.126$   \\
\hline HD, On-off with D  & HD, On-off with D   & $2.166$  \\
\hline HD, On-off with D \& S & HD, On-off with D \& S  & $2.231$  \\
\hline HD, On-off with D & On-off with D & $2.543$ \\
\hline HD, On-off with D \& S & On-off with D \& S & $2.557$ \\
\hline On-off with D & On-off with D & $2.688$ \\
\hline On-off with D \& S & On-off with D \& S & $2.782$ \\
\hline
\end{tabular}
\label{tab}
\end{table}

We have further investigated the robustness of the nonlocality-testing schemes against imperfections in detection and source preparation.
In those cases in which quantum nonlocality could be observed, we found that the single-mode squeezing operation improves the robustness with respect to both the detection efficiency, $\eta$, and the single-photon source efficiency, $p$. The improvements were relatively small for the 1 homodyne-measurement case, but quite significant for the 0 homodyne-measurement scheme. In the latter scheme, the improvements in the bounds were from $\eta \approx 0.825$ to $\eta\approx 0.78$ and $p \approx 0.83$ to $p\approx0.8$. The observed increase in violation and robustness can be explained as follows. In the 0 and 1 photon manifold, any measurement can be represented as a mixture of the identity and three Pauli operators. The on-off detection in our scheme corresponds to the $\sigma_z$ operator and the effect of displacement and squeezing is to rotate this operator in the operator space. By choosing the displacement and squeezing amplitudes carefully, one can therefore maximize the CHSH violation within the measurement setting.

The required numbers are within reach of state-of-the-art techniques. Superconducting transition-edge sensors offer single-photon detection with an efficiency as high as $95\%$ at 1,550 nm \cite{Lita:08} whereas a resonantly driven quantum dot in a micropillar can generate near-perfect single photons \cite{Ding2016}. Squeezing, the more difficult of the two operations, has also been demonstrated with a high-fidelity in Ref.~\cite{YoshikawaFurusawa2007}, based on which we expect that $r=0.24 \approx 2.1{\rm dB}$ to be implemented with more than $90\%$ fidelity. At this point, a discussion on the fidelity of the Gaussian operations is perhaps in order. We have not studied the effects of imperfections in the operations in this work, but they will certainly play an important role in actual experiments. This issue will be best addressed in more specialized works that have specific implementations in mind, but some general features can be postulated. Firstly, given that the role of these operations are to `rotate' the effective spin operators (when combined with an on-off detection), possible non-Gaussianity need not necessarily be harmful. Instead depending on actual implementations they could help enlarge the achievable effective spin-operator space and thus achieve a larger violation than those reported in this work. Secondly, non-unitarities in the operations are most likely to be detrimental as they signify departure from ideal spin measurements.



Potential future works include checking whether generalizations to many settings with binary outcomes \cite{Gisin1999} and/or two settings with ternary outcomes \cite{CollinsPopescu2002} (with 2 on-off detectors) are useful for detecting single-photon quantum nonlocality with homodyne measurements. Another interesting avenue is to check what other experimentally doable operations can be used to improve robustness against imperfections further or consider other types of imperfections such as dark-counts in on-off detectors and imperfections in Gaussian operations.

\begin{acknowledgments}
SYL and JK were supported by ICT R\&D program of MSIP/IITP
[10043464, Development of quantum repeater technology for the application to communication systems].
SYL and CN thank Se-Wan Ji for useful comments.
\end{acknowledgments}

\bibliography{mybib}

\end{document}